\documentclass[a4paper,twocolumn]{article}
\usepackage{epsfig}
\author{Gunnar Haeffler, Andreas E.~Klinkm{\"u}ller, Jonas Rangell,
 Uldis Berzinsh and  Dag Hanstorp\\{\small\it Dept.~of Physics,
G{\"o}teborg University and Chalmers University of Technology, S-412\,96
G{\"o}teborg, Sweden}}

\title{The electron affinity of tellurium}

\begin{document}
\twocolumn[
\maketitle

\begin{abstract}
The electron affinity of tellurium has been determined to
1.970\,876(7)~eV. The threshold for photodetachment of
Te$^{-}(^{2}\!P_{3/2})$ forming neutral Te in the ground state was
investigated by measuring the total photodetachment cross section
using a collinear laser-ion beam geometry. The electron affinity was
obtained from a fit to the Wigner law in the threshold region.
\end{abstract}
\begin{flushleft}
\end{flushleft}\bigskip
]

\section{Introduction}
Negative ions are fragile quantum systems, which differ considerably 
from neutral atoms and positive ions in numerous important aspects
\cite{Buc-94,Blo-95}.
Most prominently the binding energy of the outermost electron is
substantially lower than in iso-electronic atoms due to the more
efficient screening of the nuclear charge by the other electrons
\cite{Hot-85}. This
increases the significance of correlation effects for the outermost
electrons and makes negative ions a critical testing  ground for atomic
theory \cite{Sal-96,Fis-92}.

The short range potential of negative ions typically only sustains one
bound state \cite{Hot-85}. In the rare cases with more than one bound
state they are of the same parity as the ground state and
consequently inaccessible to one-photon electric dipole transitions. Hence,
the electron affinity (EA) is one of very few properties of a negative ion
that can be determined with high accuracy. 
The potentially most accurate method to determine EA  
is the so called laser photodetachment threshold (LPT) method
\cite{Neu-85} in a collinear geometry. In more recent years we have applied 
this method to Iodine 
\cite{Han-92-1} to determine the EA and to Chlorine \cite{Ber-95-4} 
where we measured both the EA and the
isotope shift between $^{35}$Cl$^{-}$ and $^{37}$Cl$^{-}$.

In this paper we investigate the EA of Tellurium.
If comparing with other elements 
with a relatively high electron affinity this quantity is
determined with a relatively large uncertainty 
\cite{Hot-85,Sla-77,Tho-96-2}. The aim of 
this work is to achieve an improvement of this value by using the 
LPT method in a collinear geometry. 

\begin{figure*}
\begin{center}
\epsfig{file=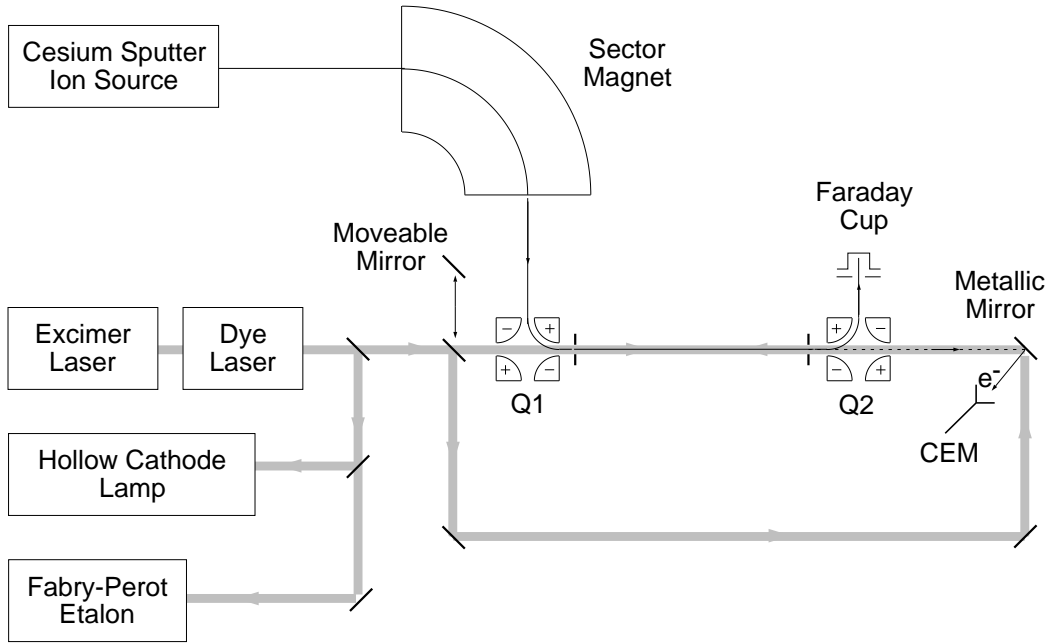, width=0.9\textwidth}
\caption{\label{fig:setup}\sloppy Experimental setup: Schematic diagram of the
collinear laser and ion beam setup. The movable mirror is used to
reverse the direction of the laser beam during a scan.}
\end{center}
\end{figure*}

\section{Experiment}

\subsection{Setup}
A collinear laser-ion beam setup, schematically shown in
Fig.~\ref{fig:setup}, was used. The setup is described in more detail
elsewhere \cite{Lju-94,Han-95}. $^{130}$Te$^{-}$ ions were produced in a
Cs-sputter ion source and accelerated to an energy of approximately 4
keV. The ions were subsequently mass-analyzed before entering the
interaction-detection chamber, which was kept under a pressure of
approximately $7 \times 10^{-9}$~mbar ($7 \times 10^{-7}$~Pa). An
electrostatic quadrupole deflector guided the ion beam into the
chamber thereby allowing a collinear laser-ion beam interaction. The
interaction region was defined by two apertures with a diameter of
$3$~mm placed $0.5$~m apart. A second electrostatic quadrupole
deflector separated ions from atoms after the interaction region and
the ion current was measured in a Faraday cup. The ion current was
typically $1$~nA. Atoms created in the interaction region continued
through the second electrostatic quadrupole deflector and were
detected using a neutral-particle detector
capable of detecting single atoms.
Only atoms arriving
during a $6$\,$\mu$s gate after the laser pulse were detected. This
time-gated detection scheme was used to suppress the background mainly
caused by collisional detachment. The detector as well as the
detection scheme have been described in detail
elsewhere \cite{Han-92-2}.

A tunable dye laser pumped by an excimer laser and operated with
Rhodamine~B was used to generate the laser light. The
pulse duration was approximately $15$~ns and the pulse energy used in the
experiment was
typically $10\mu$J. 
The frequency of the light was determined by combining Fabry-Perot
fringes and optogalvanic spectroscopy using hollow cathode lamps.

A new computer program for data acquisition and
laser control by a personal computer (PC) was developed.
The
dye laser was controlled via a general purpose interface bus (GPIB). The program 
allowed scanning of
the laser frequency with different step size and different number of laser
shots per step  within the same scan. The pulses
from a channel electron multiplier (CEM) were counted by a gated photon 
counter connected to a PC via a GPIB. The signals used for calibration were
measured by means of boxcar integrators and transfered to the computer
via A/D converters.
\subsection{Procedure}
\begin{figure}
\begin{center}
\epsfig{file=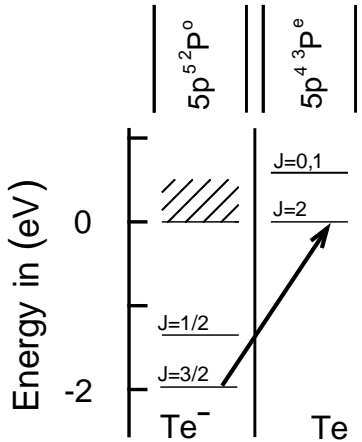, width=0.618\columnwidth}
\caption{\label{fig:Te_levels}\sloppy Excitation scheme: Selected states of
Te/Te$^{-}$. The arrow indicates the transition induced in this
experiment.}
\end{center}
\end{figure}

The electron affinity of Te corresponds to the photon energy needed to
reach the first threshold for photodetachment of the Te$^{-}$ ground
state, a process that can be denoted as
\begin{equation}
\label{eq:signal}
Te^{-}(5p^{5}\,^{2}\!P_{3/2}) + \hbar \omega \quad\rightarrow\quad 
Te(5p^{4}\,^{3}\!P_{2}) + e^{-}.
\end{equation}
The angular momentum of the
outgoing electron in the vicinity of the threshold is predominantly  
$l=0$. As seen in
Fig.~\ref{fig:Te_levels}, there will be additional contribution to
 the residual
atom yield  due to the processes
\begin{eqnarray}
\label{eq:background1}
Te^{-}(5p^{5}\,^{2}\!P_{1/2}) + \hbar \omega &\rightarrow & 
Te(5p^{4}\,^{3}\!P_{2}) + e^{-},\\\nonumber
Te^{-}(5p^{5}\,^{2}\!P_{1/2}) + \hbar \omega &\rightarrow & 
Te(5p^{4}\,^{3}\!P_{1}) + e^{-},\\
Te^{-}(5p^{5}\,^{2}\!P_{1/2}) + \hbar \omega  &\rightarrow & 
Te(5p^{4}\,^{3}\!P_{0}) + e^{-}.\nonumber
\end{eqnarray}
The contribution of these processes to the signal, however, varies
slowly with the photon energy since they are far above their thresholds.
Furthermore, these signals are relatively small  since the sputter
ion source predominantly produces ground state negative ions.
The approach in this work was
therefore to determine the electron affinity by performing a LPT
measurement around the threshold for the process described by
(\ref{eq:signal}), using a collinear laser-ion beam geometry.

The total photodetachment cross section was obtained by measuring the
neutral atom yield as a function of the laser frequency. To
establish the frequency scale, the optogalvanic signal from a hollow
cathode lamp and the fringes from a Fabry-Perot etalon were recorded
simultaneously with the neutral atom signal as the laser frequency was
tuned. The Fabry-Perot fringes served as frequency markers whereas
atomic transitions in Ne \cite{Kau-72,Cha-94} or Ar \cite{Min-73} 
from two different hollow cathode lamps
provided an absolute calibration of the scale. Within one frequency
scan only one hollow cathode lamp was used, but by performing a number
of scans with each lamp a control of the frequency calibration was
obtained.
\begin{figure}
\begin{center}
\epsfig{file=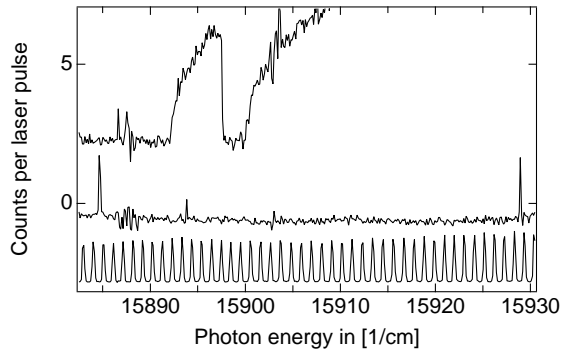, width=\columnwidth}
\protect\caption{\label{fig:scan_results}\sloppy Scan overview:
The three curves show the photodetachment signal, the reference lines 
(Ne) and the 
Fabry-Perot fringes. In the photodetachment signal, the left threshold is 
obtained with anti-parallel laser and 
ion beams and the  right threshold with parallel beams. 
The background is partly due to photodetachment of the
Te$^{-}(5p^{5}\,^{2}$P$_{1/2}$) state. The signal saturates at about 7 counts
per laser pulse.
In the vicinity of the calibration lines  and the
photodetachment thresholds the step-length was 1\,pm and 100 shots per
point were taken. 
Elsewhere the step-length was 5\,pm and 10 shots per
point were taken. The vertical scale (Counts per laser pulse) is only valid for the 
photodetachment signal.}
\end{center}
\end{figure}
The direction of the laser beam was reversed within each scan to
obtain the photodetachment cross section both for the case of parallel ion
and laser beams, where the photon energy seen by the ions is
red-shifted due to the Doppler shift,
 as well as in the case of anti-parallel beams, where a blue-shift is
obtained. Each scan covered a frequency
range wide enough to record one identified transition in Ar or Ne in
the optogalvanic spectrum below and one above the two Doppler shifted
thresholds.

A larger frequency step size and a smaller number of laser pulses were
used in the frequency ranges between the calibration lines and the
thresholds where the only relevant information was the number of
Fabry-Perot fringes to keep track of the relative frequency scale. In
this manner the scanning time was reduced and hence any long time
drifts of the system were minimized.

\section{Results}
The result of a typical measurement of the neutral atom signal,
fringes from the Fabry-Perot etalon and the signal from optogalvanic
spectroscopy, using a hollow cathode lamp is shown in
Fig.~\ref{fig:scan_results}. Two thresholds, both representing
photodetachment from the Te$^{- }(^{2}\!P_{3/2})$ ground state through
the process described in (\ref{eq:signal}), are seen in the spectrum.
At lower photon energies the laser and ion beams propagate 
anti-parallel.
The direction of the laser beam was reversed at a photon energy of about
15\,882 cm$^{-1}$ hence giving a second measurement of the now blue shifted 
threshold.  By the use of Ar transitions for the frequency calibration a
typical scanning range was $629.882$~nm to $628.022$~nm
($15876$~cm$^{-1}$ to $15923$~cm$^{-1}$) and in the case of Ne
transitions the scanning range was
$629.564$~nm to $627.746$~nm ($15884$~cm$^{-1}$ to
$15930$~cm$^{-1}$) (the used reference lines are tabulated in 
Tab.~\ref{Ref:tab}).
In regions close to optogalvanic signal
peaks and photodetachment thresholds the laser wavelength
step was 1\,pm and in between, where it only was important to keep
track of the number of Fabry-Perot fringes, the step size was
increased to 5\,pm. In the vicinity of the photodetachment threshold
100 laser pulses were used for each frequency step whereas this number
was decreased to 10 in other regions.

\begin{figure}
\begin{center}
\epsfig{file=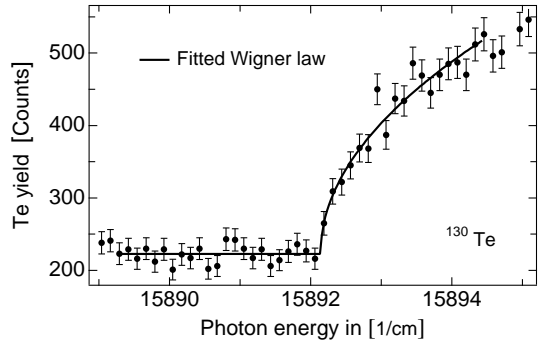, width=\columnwidth}
\protect\caption{\label{fig:threshold}\sloppy Te$(5p^{4})+\epsilon s$ threshold: 
Measurement of the relative total photodetachment cross section around the
Te$(5p^{4})+\epsilon s$ threshold with anti-parallel laser and ion
beams.
The solid line is a fit of the Wigner
law (\protect\ref{fitfunc}) to the experimental data. The error bars represent the
shot noise. Each data point was acquired during 100 laser pulses.
The background is 
mainly due to photodetachment of Te$^{-}(^{2}\!P_{1/2})$ through 
the processes described in (\protect\ref{eq:background1}).}
\end{center}
\end{figure}

To obtain the photodetachment threshold energy $E_{0}$ we fitted  the
Wigner law \cite{Wig-48} for {\it s}-wave
detachment,
\begin{equation}\label{fitfunc}
\sigma (E) =
\left\{
\begin{array}{ll}
a + b\sqrt{E-E_{th}}\;,& \quad E > E_{th}\,,\\
a\; ,& \quad E < E_{th}
\end{array}
\right.
\end{equation}
to our data by adjusting the parameters: {\it a} for the non-resonant
background, {\it b} for  the cross section amplitude and $E_{th}$ for
the threshold energy. The value of the threshold energy parameter is
either blue shifted $E^{b}_{0}$, for parallel laser and ion beams, 
or red shifted $E_{0}^{r}$,
for anti-parallel laser and ion beams 
(Fig.~\ref{fig:scan_results} and \ref{fig:threshold}). To attain a threshold
energy, $E_{0}$,  corrected for the Doppler shift to all orders the
geometric mean of the red and blue shifted threshold energy has to be
taken:
\begin{equation}
\label{eq:Doppler_elimination}
E_{0} = \sqrt{E^{b}_{0}E^{r}_{0}}.
\end{equation}
The final value, $\overline{E_{0}} = 15\,896.18(5)$~cm$^{-1}$  is 
a weighted average of eight measurements with frequency calibration obtained
from Ar transitions and ten measurements using Ne transitions. 
Average values have also been calculated individually for
the sets of $E_{0}$ values obtained by calibrating to Ne and Ar
transitions respectively. These values are presented in
Tab.~\ref{EAvalues}.

There are two major contributions to the uncertainty of $E_{0}$. There
is a statistical error of 0.01~cm$^{-1}$ corresponding to the spread of
fitted threshold values. Second, there is an 
uncertainty related to the laser intensity profile. We have estimated
this uncertainty by analyzing Fabry-Perot fringes and the atomic reference lines 
to  be less than 0.04~cm$^{-1}$, which is one fifth of the
laser frequency bandwidth.

To converted our value of $\overline{E_{0}}$
from cm$^{-1}$ to eV we used 
the recommended factor of
(1/8\,065.5410) [eV/(cm$^{-1}$)]~\cite{Coh-88} yielding the value
1.970\,876(7)~eV for the EA\@. 

\section{Discussion}
We have obtained an electron affinity of $1.970\,876(7)$~eV. The
uncertainty is of the same order of magnitude as the laser
bandwidth. We have thus shown that the background contribution from
photodetachment of Te$^{-}(^{2}\!P_{1/2})$ is not critical for the
accuracy achievable with our laser system. An important part of 
this experiment was the use of a computer program that can vary 
both the wavelength-step and the
acquisition times for each wavelength. This  
has proven to be most valuable in order to 
improve the statistics and simultaneously decrease the total acquisition time,
hence diminishing the probability of long term drifts.
The value we have achieved is of comparable accuracy as the best EA
determinations \cite{Blo-95,Hot-85}, and our 
two independent calibrations add reliability to this measurement.

This new value of the electron affinity of Tellurium is 45 times more accurate than 
the value of Slater \cite{Sla-77} and fall well within their error-bars. 
The more recent value of Th{\o}gersen 
\cite{Tho-96-2} agrees with ours within two error bars. 
\begin{table}
\begin{center}
\begin{tabular}{lcr@{.}l}
\hline\hline
Ref. line & \# of spectra & \multicolumn{2}{c}{EA in eV}\\ \hline
Neon & 10 & 1&970\,878\,0(14)\\
Ar & 8  & 1&970\,874\,6(22)\\
All & 18 & 1&970\,876\,4(13)\\
\multicolumn{2}{c}{Final} {\small (with systematic errors)} & 1&970\,876(7)\\
\hline\hline
\end{tabular}  
\protect\caption{\label{EAvalues}\sloppy EA of Te: 
The two first lines shows the values obtained using transitions in Neon and 
Argon for the wavelength calibration, and the third line shows their 
weighted average. The uncertainty given is then only the statistical 
(one sigma). In our final value the estimated systematic 
uncertainty is included as described in the text.}
\end{center}
\end{table}
\begin{table}
\begin{center}
\begin{tabular}{lr@{.}l}
\hline\hline
Element & \multicolumn{2}{c}{Line  (cm$^{-1}$)}\\ \hline
Ne \cite{Cha-94,Kau-72} & 15\,929&216(4)\\
Ne \cite{Cha-94,Kau-72} & 15\,884&396(6)\\
Ar \cite{Min-73} & 15\,922&598(5)\\
Ar \cite{Min-73} & 15\,876&508(5)\\
\hline\hline
\end{tabular}
\protect\caption{\label{Ref:tab}\sloppy Calibration lines: Transition energy of the 
calibration lines used in this experiment. The Ne transitions are
$(^{2}P_{3/2})4d[1/2]J=0\,\rightarrow\,(^{2}P_{1/2})3p[1/2]J=1$ and
$(^{2}P_{1/2})5s[1/2]J=1\,\rightarrow\,(^{2}P_{1/2})3p[3/2]J=1$, were
all involved lines are $2p^{5}nl$. We calculated the transition energies
from the tabulated interferometrically
\protect\cite{Kau-72}
or by Fourier transform spectroscopy 
\protect\cite{Cha-94} 
determined level energies.  The Ar transition energies are calculated 
from interferometrically determined levels, as presented in Tab {\sc V} of
\protect\cite{Min-73}
.}
\end{center}
\end{table}
\section{Acknowledgments}
Financial support for this research project has been obtained from the
Swedish Natural Science Council (NFR). Personal support was received
from Chalmers University of Technology for Uldis Berzinsh. R. L. Kurucz 
at Harvard-Smithsonian Center for Astrophysics is gratefully acknowledge 
for supply us with their Atomic Line List.
%

\begin{thebibliography}{10}

\bibitem{Buc-94}
S.~J. {Buckman} and C.~W. {Clark},
\newblock Rev. Mod. Phys. {\bf 66}, 539 (1994).

\bibitem{Blo-95}
C.~{Blondel},
\newblock Physica Scripta {\bf T58}, 31 (1995).

\bibitem{Hot-85}
H.~{Hotop} and W.~C. {Lineberger},
\newblock J. Phys. Chem. Ref. Data {\bf 14}, 731  (1985).

\bibitem{Sal-96}
S.~{Salomonson}, H.~{Warston}, and I.~{Lindgren},
\newblock Phys. Rev. Lett. {\bf 76}, 3092 (1996).

\bibitem{Fis-92}
C.~F. {Fischer} and T.~{Brage},
\newblock Can. J. Phys. {\bf 71}, 1283 (1992).

\bibitem{Neu-85}
D.~M. {Neumark}, K.~R. {Lykke}, T.~{Andersen}, and W.~C. {Lineberger},
\newblock Phys. Rev.~A {\bf 32}, 1890 (1985).

\bibitem{Han-92-1}
D.~{Hanstorp} and M.~{Gustafsson},
\newblock J. of Phys.~B: At., Mol. Opt. {\bf 25}, 1773  (1992).

\bibitem{Ber-95-4}
U.~{Berzinsh} et~al.,
\newblock Phys. Rev.~A {\bf 51}, 231 (1995).

\bibitem{Sla-77}
J.~{Slater} and W.~C. {Lineberger},
\newblock Phys. Rev.~A {\bf 15}, 2277 (1977).

\bibitem{Tho-96-2}
J.~{Th{\o}gersen} et~al.,
\newblock Phys. Rev.~A {\bf 53}, 3023 (1996).

\bibitem{Lju-94}
U.~{Ljungblad}, A.~{Klinkm{\"u}ller}, and D.~{Hanstorp},
\newblock A new apparatus for studies of negative ions,
\newblock in {\em Fifth european workshop on the production and application of
  light negative ions}, edited by M.~{Hopkins} and S.~{Fahy}, pages 35--40,
  Glasnevin, Dublin 9, Ireland, 1994, Dublin City University.

\bibitem{Han-95}
D.~{Hanstorp},
\newblock Nucl. Instrum. Methods Phys. Research {\bf 100}, 165 (1995).

\bibitem{Han-92-2}
D.~{Hanstorp},
\newblock Meas. Sci. Technol. {\bf 3}, 523  (1992).

\bibitem{Kau-72}
V.~{Kaufman} and L.~{Minnhagen},
\newblock J. Opt. Soc. Am. {\bf 62}, 92 (1972).

\bibitem{Cha-94}
E.~S. {Chang} and W.~G. {Schoenfeld},
\newblock Physica Scripta {\bf 49}, 26 (1994).

\bibitem{Min-73}
L.~{Minnhagen},
\newblock J. Opt. Soc. Am. {\bf 63}, 1185 (1973).

\bibitem{Wig-48}
E.~P. {Wigner},
\newblock Phys. Rev. {\bf 73}, 1002  (1948).

\bibitem{Coh-88}
E.~R. {Cohen} and B.~N. {Taylor},
\newblock J. Phys. Chem. Ref. Data {\bf 17}, 1795 (1988).

\end{thebibliography}

\end{document}